\documentclass[aps,prl,showpacs,preprintnumbers,twocolumn,superscriptaddress]{revtex4-2}
\usepackage{amsmath,amssymb}
\usepackage{bm}
\usepackage{tipa}
\usepackage{upgreek}
\usepackage{comment}
\usepackage{mathrsfs}
\usepackage{graphicx}
\usepackage{braket}
\usepackage{enumitem}
\usepackage{natbib}
\usepackage{mathbbol}
\usepackage{booktabs}
\usepackage{gensymb}
\usepackage[normalem]{ulem}
\usepackage{color}
\usepackage[colorlinks,bookmarks=true,citecolor=blue,linkcolor=red,urlcolor=blue]{hyperref}
\usepackage{hyperref}

\usepackage{pifont}

\begin{document}

	\title{Single-parameter variational wavefunctions for quantum Hall bilayers} 
	
	\author{Qi Hu}
	\affiliation{Department of Physics, ETH Zurich, Otto-Stern-Weg 1,
8093 Zurich, Switzerland}

	\author{Titus Neupert}
	\affiliation{Department of Physics, University of Zurich, Winterthurerstrasse 190, 8057 Zurich, Switzerland}
 
	\author{Glenn Wagner}
	\affiliation{Department of Physics, University of Zurich, Winterthurerstrasse 190, 8057 Zurich, Switzerland}

	\begin{abstract}
    Bilayer quantum Hall states have been shown to be described by a BCS-paired state of composite fermions. However, finding a qualitatively accurate model state valid across all values of the bilayer separation is challenging. Here, we introduce two variational wavefunctions, each with a \emph{single} variational parameter, which can be thought of as a proxy for the BCS order parameter. Studying systems of up to $9+9$ electrons in a spherical geometry using Monte Carlo methods, we show that the ground state can be accurately described by these single-parameter variational states. In addition, for the first time we provide a numerically exact wavefunction for the Halperin-111 state in terms of composite fermions.
	\end{abstract}
	
 	\maketitle

\textit{\textbf{Introduction}} --- Non-Fermi liquids have been attracting an increasing amount of interest as a gapless phase of matter that challenges the paradigm of Fermi liquid theory \cite{NFL}. One example of a non-Fermi liquid is the compressible liquid arising at half-filling of a Landau level (LL) in the quantum Hall effect. Quantum Hall bilayers offer a platform for studying pairing instabilities of this non-Fermi liquid. 

The quantum Hall effect arises when electrons confined to two dimensions are subjected to a strong magnetic field. For particular values of the filling factor $\nu=N_e/N_\phi$, where $N_e$ is the number of electrons and $N_\phi$ is the number of magnetic flux quanta, a gapped state with a quantized Hall response is observed. The first experiments \cite{IQH_1980} observed quantized Hall plateaus at integer $\nu$, i.e.~the integer quantum Hall effect (IQHE). However, soon thereafter further plateaus at fractional $\nu$ were observed \cite{FQH_1982} marking the discovery of the fractional quantum Hall effect (FQHE). Whereas the IQHE can be described effectively as a band insulator of non-interacting electrons, the FQHE fundamentally requires electron-electron interactions in order to open up a gap. 

One of the most successful approaches to the quantum Hall effect consists of thinking in terms of composite fermions (CFs) --- composite objects of electrons bound to an even number of flux quanta \cite{CompositeFermionsJain,CompositeFermionsHeinonen}, since this allows one to bridge the gap between the IQHE and FQHE. At the mean field level, the CFs experience a different effective magnetic field than the electrons such that integer values of their effective filling factor $\nu_\textrm{CF}$ correspond to fractional values of $\nu$. This unifying framework allows one to describe the fractional quantum Hall effect of strongly interacting electrons as an integer quantum Hall effect of weakly interacting CFs. Furthermore, the compressible state observed at $\nu=1/2$ \cite{CFL_exp} can be viewed as a CF Fermi liquid (CFL). However, the CFs experience a fluctuating gauge field which can lead to non-Fermi liquid behaviour \cite{HLR} and the residual interactions between the CFs are still able to generate instabilities such as pairing instabilities and open up a gap. Paired states of composite fermions such as the Moore-Read state are indeed candidates for the elusive gapped $\nu=5/2$ state~\cite{MOORE1991362,Willett_2013}. 

Another platform to study pairing of composite fermions is a quantum Hall bilayer with total filling factor $\nu=1$. The electrons are confined to two layers with layer separation $d$, with each layer at half filling $\nu=1/2$. The typical distance between electrons in the same layer is given by the magnetic length $\ell_B = \sqrt{\hbar/e B}$ and therefore the ratio of interlayer to intralayer interaction strength is roughly $\frac{1}{d/\ell_B}$. By tuning the ratio $d/\ell_B$, the two competing interactions  can be tuned. At large $d/\ell_B$ the composite fermions form two decoupled composite Fermi liquids in the two layers \cite{HLR}, for which numerically exact wavefunctions can be written. At small $d/\ell_B$, electron-hole pairs form an exciton condensate, the so-called 111-state~\cite{Halperin111,Experiment_Review}. The limits $d\to 0$ and $d\to\infty$ of the quantum Hall bilayer are thus well-understood. However, one difficulty is that the two limits are described in terms of different quasiparticles (electrons at  $d\to 0$ vs.~CFs at $d\to \infty$). Much theoretical work has been devoted to understanding the nature of the state at intermediate distances and the connection between these two well-understood limits \cite{Moon_Review,Eisenstein2004,Bonesteel,p_wave,Pwave2,Ezawa_2009,ShouCheng,HF1,HF2,HF3,HF4,HF5,HF6,ED1,ED3,ED0,DMRG,Park1,Park2,Simon1,Simon2,Simon3,Kimchi,Sodemann,Milovanovic,Ye1,Ye2,ICCFL,Cipri_thesis,CipriBonesteel,papicThesis,Bosonization1,Bosonization4}.

Recently, it has been proposed that at intermediate distances the composite fermions in a quantum Hall bilayer pair up in a BCS-like fashion and undergo a BEC-BCS crossover, from a BCS-like state at large $d/\ell_B$ to a BEC-like state at small $d/\ell_B$ \cite{Crossover,Halperin2020}. Experiments on double layers of graphene have shown that as $d/\ell_B$ decreases, one goes from a regime where the pairing temperature and the condensation temperature coincide (BCS regime) to a regime where the pairing temperature lies significantly above the condensation temperature (BEC regime), as expected for the BEC-BCS crossover \cite{Crossover}. Besides this experimental evidence, exact diagonalization results show that a $s$-wave BCS trial state with CFs in one layer paired with anti-CFs in the other layer has high overlaps with the exact ground state \cite{Wagner2021} for any interlayer separation. An Eliashberg calculation of the pairing of CFs and anti-CFs mediated by the fluctuating gauge field they experience indeed finds a dominant $s$-wave pairing channel \cite{ruegg2023pairing}.

In the present work, we study trial wavefunctions for the BEC-BCS crossover in quantum Hall bilayers. In contrast to previous work, we use trial wavefunctions with a \emph{single} variational parameter. Given that the Hilbert space size is exponentially increasing with the number of electrons, the fact we can capture the ground state with a single variational parameter shows that the wavefunction describes the correct physics. Moreover, our trial state captures the Halperin-111 state, which is known to be the exact ground state of the quantum Hall bilayer at $d=0$. The 111 state is usually understood as a condensate of interlayer electron/hole excitons. However, we show that it can also be represented as a condensate of CF/anti-CF excitons up to numerical precision. Therefore,  we find that the quantum Hall bilayer at $\nu=1/2+1/2$ can be \emph{entirely} described in terms of CFs irrespective of $d$.

\textit{\textbf{Methods}} --- The starting point for our analysis is a trial wave function for the quantum Hall bilayer introduced in Ref.~\cite{Wagner2021}. This wavefunction describes $s$-wave BCS pairing of CFs in one layer with anti-CFs in the other layer. The CFs are organized in CF LLs (``$\Lambda$-levels"). For each of these $\Lambda$-levels a separate pairing parameter $g_n$ is introduced, where $n=0,1,...,N_\Lambda$ is the $\Lambda$-level index. $N_\Lambda$ is the maximum $\Lambda$-level that is included and in the following we will always set $N_\Lambda=N_1-1$ which is required to capture the 111 state as explained below. The trial wavefunction written as appropriate for the spherical geometry that we use here is \cite{Wagner2021}
\begin{eqnarray}
    \Psi_{\textrm{BCS}} &=& \prod_{i<j}(\Omega_i-\Omega_j)^2(\varpi_i-\varpi_j)^{*2}\det(G)  \nonumber \\
    G(\Omega_i, \varpi_j)  &=& \sum_{n,m} g_n \,  \tilde Y_{q,n,m}(\Omega_i)  \tilde Y^*_{q,n,m}(\varpi_j),  \label{eq:trialwfmaintext}
\end{eqnarray}
where $\Omega_j=(\theta_j,\varphi_j)$ is the spinor coordinate of the $j$-th electron in the top layer, $\varpi_i$ is the spinor coordinate of the $i$-th hole in the bottom layer and the notation $(\Omega_i-\Omega_j)$ is shorthand notation for a Jastrow factor. $\tilde Y_{q,n,m}$ are the Jain-Kamilla projected monopole harmonics \cite{Kamilla_Jain}, $2q$ is the net flux experienced by the CFs and $m$ is the $L_z$ angular momentum quantum number. We consider the case of a balanced bilayer with $N_1$ electrons per layer and a total number of flux quanta $N_\phi=2N_1-1$, which corresponds to a filling factor $\nu=1/2+1/2$ in the thermodynamic limit, with a shift appropriate for observing the CFL in each individual layer when they are decoupled. 

In Ref.~\cite{Wagner2021}, the number of $\Lambda$-levels that are included---and hence the number of variational parameters---is proportional to the system size $N_1$: $N_\Lambda=N_1-1$.
In the present work, we use the same trial wavefunction~\eqref{eq:trialwfmaintext}, however we use an ansatz for the parameters $g_n$ such that there is only one variational parameter. We use two different types of ansatz: (i) We use the BCS order parameter $\Delta$ as the variational parameter. The BCS prediction for the occupation probability of the composite fermion orbitals with index $n$ and energy $\varepsilon_n$ is
\begin{equation}
   p_n=\frac{1}{2}\bigg(1-\frac{\varepsilon_n}{\sqrt{\varepsilon_n^2+\Delta^2}}\bigg) 
   \label{eq:occ}
\end{equation}
and by solving the inverse problem we may deduce the parameters $g_n$ corresponding to a given $\Delta$. We measure $\varepsilon_n$ in units of the Fermi energy such that $\Delta$ is dimensionless. Note that the $\Lambda$-level index $n$ of the CFs can be thought of as a momentum $\mathbf{k}$, which allows one to compute $\varepsilon_n$ (see Supplementary Material Eq.~\eqref{eq:epsilon_n}). (ii) We use a parameter $\alpha$ as the variational parameter such that \begin{equation}
    g_n=e^{\alpha n}.
    \label{eq: alpha-ansatz}
\end{equation}

In order to study larger systems than are accessible with exact diagonalization, we use Monte-Carlo methods to minimize variational energies. In Ref.~\cite{Wagner2021} it was shown that for systems of up to $7+7$ electrons, the BCS trial state with $N_\Lambda=N_1-1$ always has at least 0.95 overlap squared with the exact diagonalization ground state. In the present work, therefore, the same number of variational parameters are used for energy minimization and the resulting BCS state is chosen as the reference ground state.
Using the fact that having large overlaps is a transitive feature, we may deduce that the single-parameter optimized variational state has high overlaps with the exact ground state (provided it has high overlap with the reference state). 

Since the optimization at large interlayer separation is trivial (the BCS trial state exactly reduces to the CFL state in a certain limit), whereas at small interlayer separation the optimization can have difficulties converging, we pick the 111-state for the importance sampling of the Monte-Carlo samples.

\begin{figure}[t]
    \centering
    \includegraphics[width=1.0\linewidth]{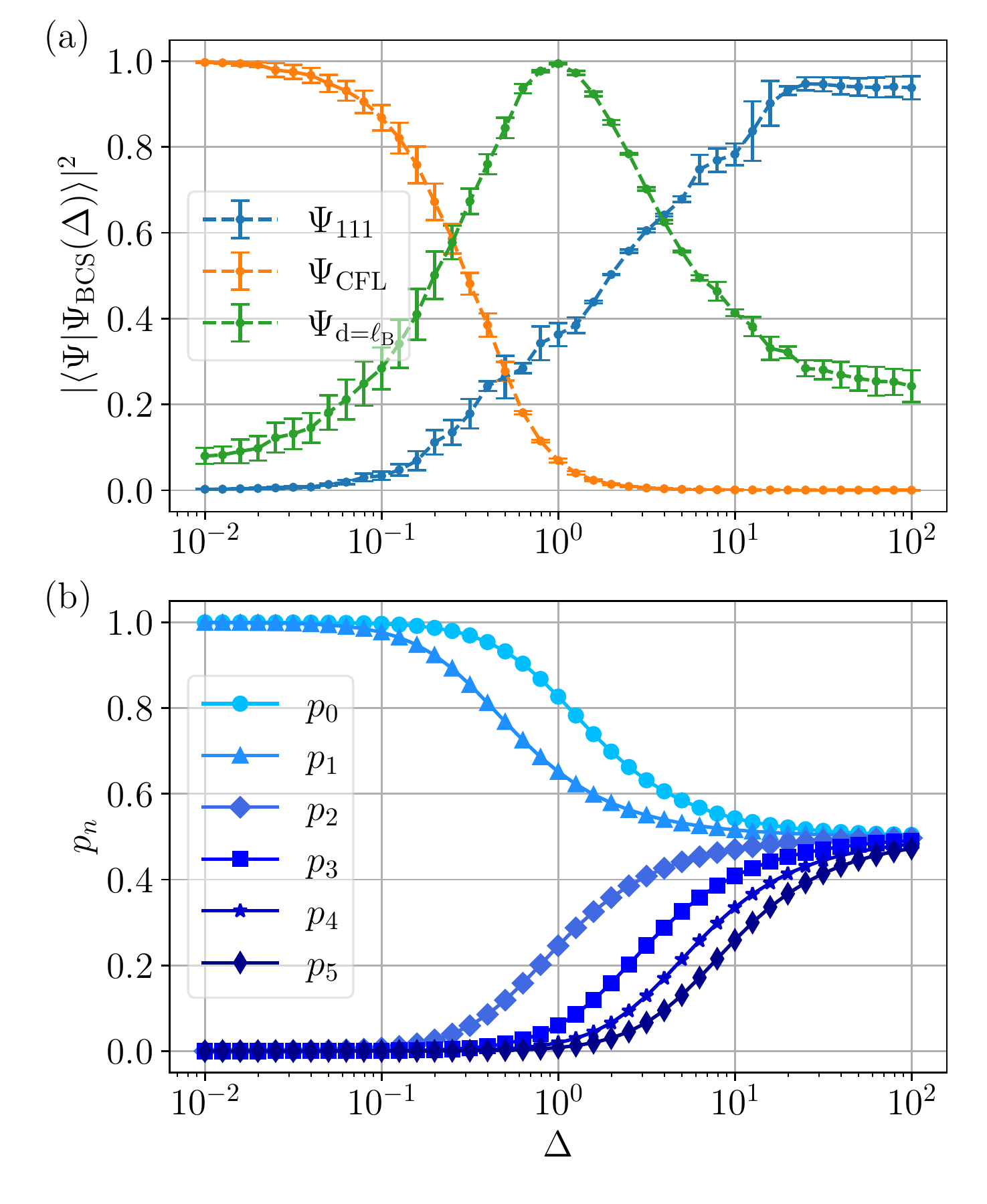}
    \caption{
    \textbf{Variational results for $\Delta$-optimization}
        (a) Overlaps with three representative states (the two model states and the (energy-optimized) ground state at $d / \ell_B = 1$) for $-2 \le \log_{10} (\Delta) \le 2$ in a 6+6 system. The CFL is described by the limit $\Delta\to0$ (BCS regime), while the 111 state is well-described by the limit $\Delta\to\infty$ (BEC regime). At intermediate distances $d\sim\ell_B$ we have the best overlap with a state having $\Delta\sim1$. (b) Orbital occupation probability for variational states. Occupation probabilities are analytical results evaluated according to Eq.~\eqref{eq:occ}. In the limit $\Delta\to0$ only the lowest two CF shells are filled, while at $\Delta\to\infty$ all CF shells have equal occupation.
    }
     \label{fig:Delta_opt}
\end{figure}

\textit{\textbf{Results for $\Delta$-optimization}} --- We first attempt to use the BCS order parameter $\Delta$ as a variational parameter. From a given set of BCS coupling constants $g_n$ we can extract the CF orbital occupation numbers via the prescription outlined in Ref.~\cite{Simon1}. These are related via Eq.~\eqref{eq:occ} to $\Delta$. Since the wavefunction is written in terms of $g_n$, to evaluate the wavefunction for a given $\Delta$ we need to first solve an optimization problem to find the corresponding $g_n$. We can then optimize $\Delta$ to find the lowest energy configuration. Since we have two nested variational problems, this is a computationally intensive method, which motivates us to later investigate a different ansatz which directly gives the $g_n$ coefficients. 

As shown in Fig.~\ref{fig:Delta_opt}, for $\Delta\to0$ we recover the CFL wavefunction which has extremely high overlaps ($>99.9\%$ for $N_1=6$) with the exact diagonalization ground state at $d\to\infty$. For $\Delta\to\infty$ we recover a state that has very high overlap with the 111-state, which is consistent with the picture from the Chern-Simons theory of this trial state \cite{Crossover}: For tightly-bound CF/anti-CF pairs, the fluxes attached to the CF and anti-CF cancel, making this CF/anti-CF-exciton equivalent to an electron/hole exciton, whose condensation leads to the 111 state. For intermediate distances $d\sim\ell_B$ we find $\Delta\sim1$.

By minimizing the energy as a function of $\Delta$, we find $\Delta\propto d^{-3.4}$ scaling for $d\gtrsim\ell_B$. This is consistent with the BEC-BCS crossover picture where $\Delta$ increases as we approach the small-$d$ BEC limit. An RG calculation for quantum Hall bilayers predicts $\Delta\propto d^{-2}$ \cite{Sodemann}, however this was derived for pairing of CFs with CFs whereas we are considering pairing of CFs with anti-CFs.

\begin{figure}[t]
    \centering
    \includegraphics[width=1.0\linewidth]{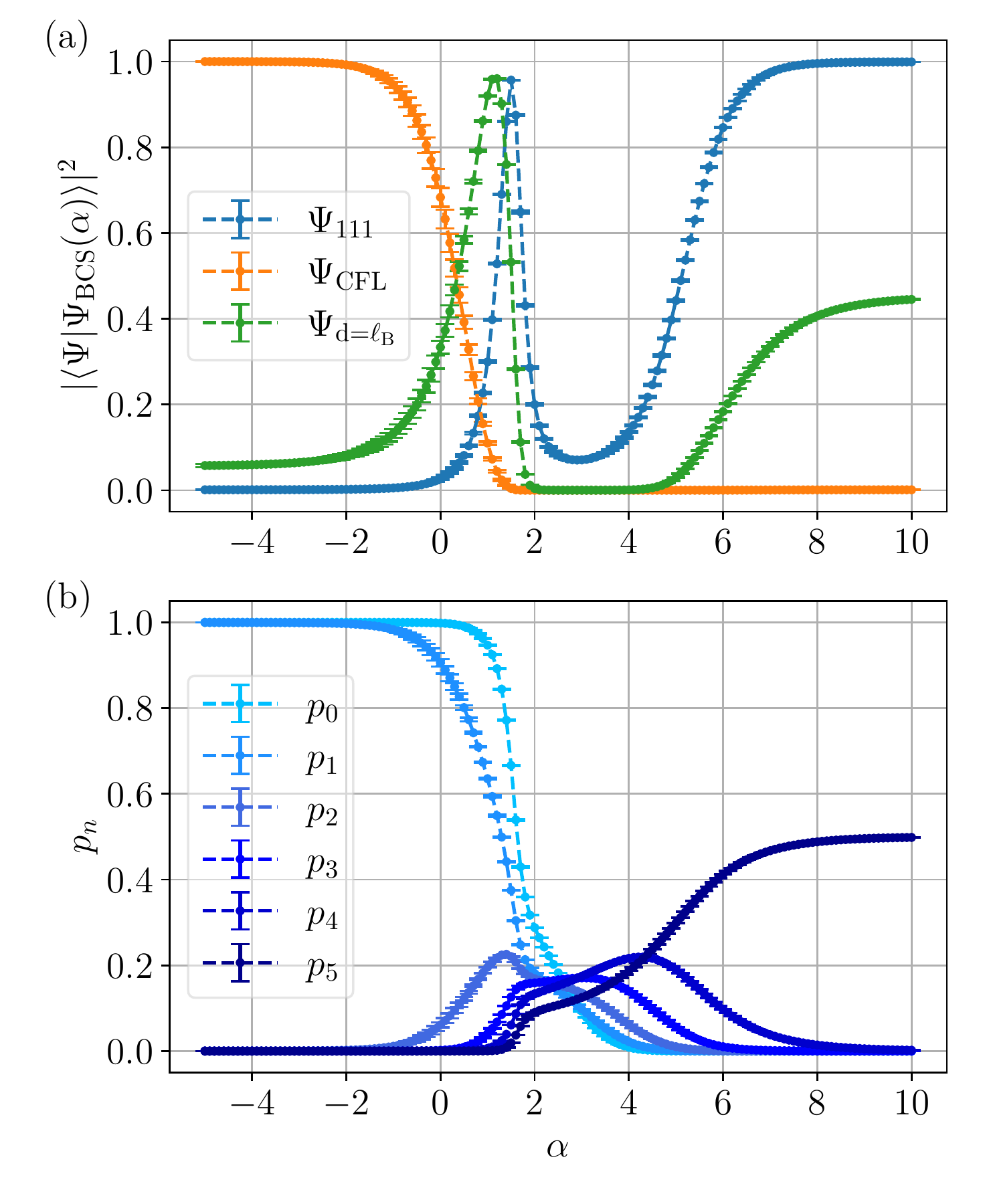}
    \caption{
    \textbf{Variational results for $\alpha$-optimization}
        (a) Overlaps with three representative states (the two model states and the (energy-optimized) ground state at $d / \ell_B = 1$) for $-5 \le \alpha \le 10$ in a 6+6 system. The CFL state has maximum overlap with the $\alpha\to-\infty$ state, the 111 state has maximum overlap with the $\alpha\to\infty$ state. At intermediate $d\sim\ell_B$, the optimum value is $\alpha\sim 1$. (b) Orbital occupation probability for variational states. For $\alpha\sim1$ we have roughly equal occupation of all CF orbitals (corresponding to $\Delta\to\infty$). The overlap with the 111 state has a local maximum at this value. In the limit $\alpha\to\infty$ which corresponds to the 111 state, only the highest CF orbital is occupied. 
    }
     \label{fig:alpha_opt}
\end{figure}

\textit{\textbf{Results for $\alpha$-optimization}} --- We now turn to a computationally more manageable approach, namely the ansatz $g_n=e^{\alpha n}$. This ansatz is motivated by the fact that when variationally optimizing the $g_n$ parameters at small $d$, they show an exponential dependence on $n$ (see Supplementary Material Fig.~\ref{fig:g_opt_arr}).

\begin{figure}[t]
    \centering
    \includegraphics[width=1.0\linewidth]{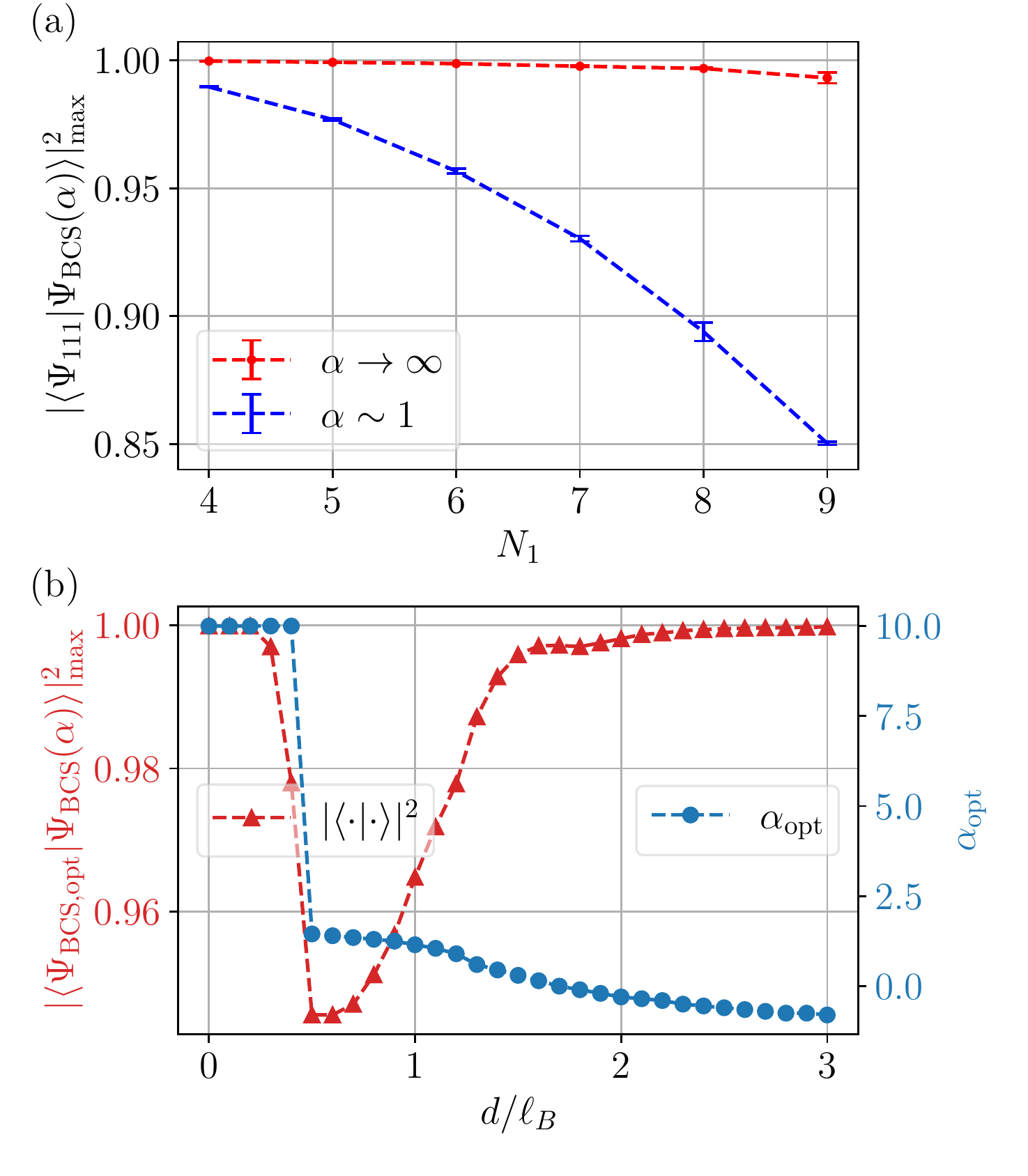}
    \caption{\textbf{Overlaps of the $\alpha$ trial state for different system sizes and interlayer separations.}
        The maximum overlap of  $\alpha$-ansatz trial state Eq.~\eqref{eq: alpha-ansatz} (a) with $\Psi_{111}$ for different system sizes $N_1$ and (b) for different interlayer distances $d$ for $6+6$ electrons. The one-parameter ansatz captures the 111 state accurately for all system sizes shown and captures the state at intermediate distances $d$ well too. We also show the optimum value of $\alpha$ as a function of interlayer separation. The optimum $\alpha$ value increases as $d$ decreases and jumps discontinuously to its maximum value in the optimization range ($\alpha=10$) around $d\sim0.5\ell_B$.
    }
     \label{fig:fig3}
\end{figure}

As shown in Fig.~\ref{fig:alpha_opt}, the limit $\alpha\to-\infty$ leads to only the lowest CF orbitals being occupied, which again reduces to the CFL wavefunction. The regime $\alpha\sim1$ has high overlaps with the 111 state. This regime leads to occupation numbers that are almost constant as a function of $n$. This can be understood as a consequence of the lowest LL (LLL) projection of the CF orbitals: those with large $n$ have small weight in the LLL and therefore need exponentially large coefficients $g_n$~\cite{Kamilla_Jain}. Furthermore, this corresponds to the regime $\Delta\to\infty$ that was previously identified as having a large overlap with the 111 state. However, we find that increasing $\alpha$ even further leads to a state with only the highest CF shell being occupied which has even better overlap with the 111 state. States in this regime are far outside the Hilbert space captured by the variational ansatz with $\Delta$ where the lowest CF shells are always occupied. We note that there is a discontinuous change in the optimum value of $\alpha$ as a function of $d$ as seen in Fig.~\ref{fig:fig3}(b).

In the limit $\alpha\to\infty$, we find a wavefunction that almost exactly reproduces the 111 state within the numerical accuracy for all system sizes up to $9+9$ particles. As shown in Fig.~\ref{fig:fig3}(a), the overlaps with the 111 state are better than 0.993 for all system sizes up to and including $9+9$. The 111 state is thus described by the wavefunction Eq.~\eqref{eq:trialwfmaintext} with $g_n=\delta_{n,N_1-1}$. This is the first time a wavefunction for the 111 state in terms of composite fermions has been written down. The CFs in each layer half fill the $\Lambda$-level with $n=N_1-1$ and there is $s$-wave pairing between the CFs in one layer and the anti-CFs in the other layer. In the 111 state description in terms of electrons and holes, electrons half fill the LLL in each layer and there is $s$-wave pairing between the electrons in one layer and the holes in the other layer. The correspondence between the two descriptions makes sense intuitively: Eq.~\eqref{eq:trialwfmaintext} describes $s$-wave pairing of CFs and anti-CFs. If we have tightly-bound CF/anti-CF pairs, then the set of coordinates $\{\Omega_i\}$ of the CFs coincides with the set of coordinates $\{\varpi_i\}$ of the anti-CFs and therefore the Jastrow factors in Eq.~\eqref{eq:trialwfmaintext} cancel. The Jastrow factors describe the flux attachment procedure and removing the Jastrow factors reduces the pairing of CFs and anti-CFs to that of electrons and holes
--- which is precisely the 111 state. For system sizes of $10+10$ electrons and above, the overlap of the $\alpha\to\infty$ state with the 111 state becomes small (see Supplementary Material Fig.~\ref{fig:model_ovlps_arr}(f)).
However in that case we are dealing with orbitals with high LL index and we caution that in that case the approximate LLL projection we use may not be accurate. Furthermore, the evaluation of CF orbitals with high LL index may suffer from numerical precision issues~\cite{Davenport}. 

In Fig.~\ref{fig:fig3}(a) we also show the overlap of the 111 state with the state with $\alpha\sim1$ as a function of system size $N_1$. As can be seen from Fig.~\ref{fig:alpha_opt}, this is a local maximum of the overlap. However, we can see that this state performs significantly worse, when the system size is increased. 

As shown in Fig.~\ref{fig:fig3}(b), the trial state with a single variational parameter captures the entire crossover from large to small $d$ very well. The limits of large and small $d$ are captured exactly to within numerical precision, while at intermediate distances, the overlap squared is always better than 0.94 for a system of $6+6$ electrons.

\textit{\textbf{Conclusion}} --- We have investigated a BCS trial wavefunction for quantum Hall bilayers which consists of pairing CFs in one layer with anti-CFs in the opposite layer. Previous work \cite{Wagner2021} used trial wavefunctions with the number of variational parameters growing proportional to the system size. Here, in contrast, we achieve high overlaps squared of better than 0.94 for up to $6+6$ electrons with a single variational parameter. Overall these are extremely high overlaps considering we are only using a single variational parameter. 

In particular, we show that for a particular choice of variational parameters, the BCS trial state which is entirely written in terms of CF orbitals has unity overlap with the 111 state (within numerical precision). The 111 state is known to be the exact ground state of the quantum Hall bilayer system at $d=0$, however it is usually written in terms of electrons. For the first time we provide the expression for the 111 state in terms of composite fermions. 

One of the interesting features of the quantum Hall bilayer system is that the large $d$ physics is most simply understood in terms of CFs, while the small $d$ physics is most simply understood in terms of electron-hole excitons. We have now shown, that CFs offer an accurate description of the system for all $d$. It would be very interesting to confirm experimentally that this is the case. Experimental evidence for a composite fermion description can come from geometric resonance experiments, as have been performed on the $\nu=1/2$ system \cite{Kang1993} and more recently on the $\nu=5/2$ system \cite{Hossain2018}. Geometric resonance experiments on quantum Hall bilayers have indeed revealed the presence of CFs \cite{Mueed}. It would be fascinating to perform such experiments on a quantum Hall bilayer as a function of the interlayer separation $d/\ell_B$ to confirm at which interlayer separation (if any) signatures of CFs disappear.  

Recent work has shown that imbalanced bilayers at filling $\nu=1/3+2/3$ also undergo a continuous transition \cite{zhang2023xy} and it would be interesting to investigate the trial wavefunctions for that scenario. We leave that to future work.

\textit{\textbf{Acknowledgements}} --- GW and TN acknowledge funding from the European Research Council (ERC) under the European Union’s Horizon 2020 research and innovation programm (ERC-StG-Neupert-757867-PARATOP). GW would like to thank Dung Nguyen, Steven Simon and Bertrand Halperin for a related collaboration on quantum Hall bilayers. The numerical simulations were performed on the Euler cluster operated by the High Performance Computing group at ETH Z\" urich.

 	\bibliographystyle{unsrtnat}
 	\bibliography{bib.bib}

\clearpage 
\newpage

\onecolumngrid
	\begin{center}
		\textbf{\large --- Supplementary Material ---\\ Single-parameter variational wavefunctions for quantum Hall bilayers}\\
		\medskip
		\text{Qi Hu, Titus Neupert, Glenn Wagner}
	\end{center}
	
		\setcounter{equation}{0}
	\setcounter{figure}{0}
	\setcounter{table}{0}
	\setcounter{page}{1}
	\makeatletter
	\renewcommand{\theequation}{S\arabic{equation}}
	\renewcommand{\thefigure}{S\arabic{figure}}
	\renewcommand{\bibnumfmt}[1]{[S#1]}
\begin{appendix}

\section{Composite fermion wavefunctions}
Consider quantum Hall bilayers on a sphere with the magnetic field normal to its surface. The single electron eigenstates are the monopole harmonics $Y_{q,n,m}(\Omega_j)$~\cite{Kamilla_Jain}
\begin{align}
    Y_{q,n,m}(\Omega_j) = &N_{qnm} (-1)^{q+n-m} e^{iq\varphi_j} u_j^{q+m} v_j^{q-m} \\
    &\times \sum^{n}_{s=0} (-1)^s \binom{n}{s} \binom{2q+n}{q+n-m-s} (v_j^* v_j)^{n-s} (u_j^* u_j)^{s},
\end{align}
where the spinor coordinates are
\begin{align} \label{eq:spinor_coordinates}
    u_j &= \cos (\theta_j/2) e^{-i \varphi_j /2} \\
    v_j &= \sin (\theta_j/2) e^{i \varphi_j /2}.
\end{align}
$q$ can be integer or half-integer and is defined via the total flux through the sphere, $N_{\phi} = 2q$. $n \in \{ 0,1,2,\dots \}$ is the LL index with 0 labeling the LLL. $m \in \{ -q-n, -q-n+1, \dots, q+n \}$ are labels for the $2(q+n)+1$ degenerate states within each LL. $\Omega_j = (\theta_j, \varphi_j)$ are the polar coordinates of electron $j$ on the sphere. $\binom{n}{k}$ is the binomial coefficient. Normalization constant $N_{qnm}$ is given by
\begin{align}
    N_{qnm} = \sqrt{ \frac{2q+2n+1}{4\pi} \frac{(q+n-m)!(q+n+m)!}{n!(2q+n)!} }
    = \sqrt{ \frac{2l+1}{4\pi} \frac{(l-m)!(l+m)!}{n!(l+q)!} }.
\end{align}
Assume there are $N_1$ electrons in each layer. We attach to each electron two flux quanta and the resulting CFs experience reduced effective flux $N_{\phi}^{\mathrm{eff}} = 2Q$ with $Q = q - (N_1 - 1)$. We can now write down the CFL state for a single layer as a Slater determinant of CFs that fill the appropriate orbitals according to Hund's rule~\cite{Hunds_rule}
\begin{align}
    \Psi_{\mathrm{CFL}}(\{ \Omega \})
    &= \mathcal{P}_{\mathrm{LLL}} \Bigl[ \mathcal{J}(\{ \Omega \}) \det \bigl[ Y_i(\Omega_j) \bigr] \Bigr] \\
    &\approx \mathcal{J}(\{ \Omega \}) \Bigl[ \det \bigl[ \mathcal{P}_{\mathrm{LLL}} Y_i(\Omega_j) \bigr] \Bigr] \\
    &\equiv \mathcal{J}(\{ \Omega \}) \det \bigl[ \tilde{Y}_i(\Omega_j) \bigr],
\end{align}
with $\mathcal{J}(\{ \Omega \})$ the Jastrow factor
\begin{align} \label{eq:Jastrow_factor_spinor}
    \mathcal{J}(\{\Omega\})
    = \prod_{j \neq k} (u_j v_k - v_j u_k)^2 e^{i (\varphi_j + \varphi_k)}
    \equiv \prod_{j<k} (\Omega_j - \Omega_k)^2,
\end{align}
where $i$ is short-hand notation for the indices $(Q, n, m)$ of filled states. We have also introduced in the last expression the short-hand notation for the Jastrow factor. $\mathcal{P}_{\mathrm{LLL}}$, proposed by Jain and Kamilla~\cite{Kamilla_Jain}, projects single particle states to the LLL, and the resulting wavefunctions are defined via $\tilde{Y}_i (\Omega_j) J_j = \mathcal{P}_{\mathrm{LLL}}[Y_i (\Omega_j) J_j]$. The explicit form of these projected states are~\cite{Kamilla_Jain}
\begin{align} \label{eq:Y_tilde}
    \tilde{Y}_{Q,n,m} (\Omega_j) = &N_{Qnm} (-1)^{Q+n-m} \frac{(2q+1)!}{(2q+n+1)!} e^{iQ\varphi_j} u_j^{Q+m} v_j^{Q-m} \\
    &\times \sum^{n}_{s=0} (-1)^s \binom{n}{s} \binom{2Q+n}{Q+n-m-s} v_j^{n-s} u_j^{s} R_j^{s, n-s},
\end{align}
where
\begin{align}
    R_j^{s, n-s} = \mathbf{U}_j^s \mathbf{V}_j^{n-s} 1
\end{align}
with
\begin{align}
    \mathbf{U}_j &= \sum_{k \neq j} \frac{v_k}{u_j v_k - v_j u_k} + \frac{\partial}{\partial u_j} \\
    \mathbf{V}_j &= \sum_{k \neq j} \frac{-u_k}{u_j v_k - v_j u_k} + \frac{\partial}{\partial v_j}.
\end{align}
The (LLL-projected) CFL state can then be written as
\begin{align} \label{eq:Psi_CFL_single-layer}
    \tilde{\Psi}_{\mathrm{CFL}}(\{ \Omega \}) = \prod_{k < l} (\Omega_k - \Omega_l)^2 \det \Bigl[\tilde{Y_i}(\Omega_j) \Bigr].
\end{align}

\section{The $s$-wave paired BCS state}
We work with an even number of total particle number $N$ and consider the balanced case, $N_1 = N_{\uparrow} = N_{\downarrow} = \frac{N}{2}$, or written in electron densities as $n_{\uparrow} = n_{\downarrow}$. Electron coordinates in top and bottom layers are labeled as $\Omega^{\uparrow}_i$ and $\Omega^{\downarrow}_i$, respectively. Since we assume the magnetic field passes through both layers simultaneously, it follows that they experience the same amount of flux. Thus, the net flux felt by the CFs in both layers after flux attachment are given by $Q_{\uparrow, \downarrow} = q - (N_{\uparrow, \downarrow} - 1) = 1/2$, where $2q = N_{\phi} = 2 N_1 - 1$. In other words, we have $\nu_{\uparrow} = \nu_{\downarrow} = 1/2$ with total filling factor $\nu_{T} = \nu_{\uparrow} + \nu_{\downarrow}$ fixed to 1.
In order to write down the $s$-wave paired BCS state, we need to perform a particle-hole transformation~\cite{PH} on the bottom layer and write their coordinates as $\varpi^{\downarrow}_{1}, \dots, \varpi^{\downarrow}_{N_{1}}$. The anti-CFs, which are holes with flux quanta attached, experience the same flux $Q = 1/2$ as the CFs in the top layer. The paired wavefunction after the LLL-projection can then be written as
\begin{equation} \label{eq:s-wave_BCS}
    \tilde{\Psi}_{\mathrm{BCS}, s}(\{\Omega^{\uparrow}\}, \{\varpi^{\downarrow}\}) = \prod_{i<j} \Bigl[ (\Omega^{\uparrow}_i - \Omega^{\uparrow}_j)^2 (\varpi^{\downarrow}_i - \varpi^{\downarrow}_j)^{*2} \Bigr] \det \Bigl[ \tilde{G} (\Omega^{\uparrow}_{i}, \varpi^{\downarrow}_{j}) \Bigr]
\end{equation}
with
\begin{equation}
    \tilde{G} (\Omega^{\uparrow}_{i}, \varpi^{\downarrow}_{j}) = \sum_{n=0}^{N_{\mathrm{LL}} - 1} \sum_{m=-(n+1/2)}^{n+1/2} g_n \tilde{Y}_{\frac{1}{2}, n, m} (\Omega_i^{\uparrow}) \tilde{Y}_{\frac{1}{2}, n, m}^* (\varpi_j^{\downarrow}),
\end{equation}
where $N_{\mathrm{LL}} = \vert \{ g_n \} \vert$ is the number of variational parameters. In practice the variational parameters $\{ g_n \}$ are always chosen to be real.

\section{111 state in terms of CFs}
The 111 state written in electron and hole coordinates is 
\begin{equation}
\label{eq:111_eh}
\Psi_{111}(\{\Omega^\uparrow\},\{\varpi^\downarrow\})=\textrm{det}\bigg[\sum_{m=-q}^{q}Y_{q,0,m}(\Omega^\uparrow_i)Y_{q,0,m}^*(\varpi^\downarrow_j)\bigg],
\end{equation}
where $q=N_1-\frac{1}{2}$. Now if we choose to fill only the $\Lambda$-level with $n=q-\frac{1}{2}$ of the CFs, then the BCS trial state is 
\begin{align}\label{eq:s_wave}
\tilde \Psi_{\textrm{BCS},s}(\{\Omega^\uparrow\},\{\varpi^\downarrow\})&=\mathrm{det}[ Y_{\tilde q,0,m_i}(\Omega_j)]^2\mathrm{det}[ Y_{\tilde q,0,m_i}^*(\varpi_j)]^2 \mathrm{det} \bigg[ \sum_{m=-q}^{q} \tilde{Y}_{\frac{1}{2},q-\frac{1}{2}, m}\left(\Omega_{i}^{\uparrow}\right) \tilde{Y}^*_{\frac{1}{2}, q-\frac{1}{2},m}\left(\varpi_{j}^{\downarrow}\right)\bigg]\\
&= \mathrm{det} \bigg[ \sum_{m=-q}^{q} J(\Omega_{i}^{\uparrow})\tilde{Y}_{\frac{1}{2},q-\frac{1}{2}, m}(\Omega_{i}^{\uparrow}) J(\varpi_{j}^{\downarrow})\tilde{Y}^*_{\frac{1}{2}, q-\frac{1}{2},m}(\varpi_{j}^{\downarrow})\bigg]
\end{align}
where $\tilde q=\frac{N_1-1}{2}$. We find that this state is a numerically accurate representation of the 111 state in terms of CFs.

\section{BCS parameters}
In the second quantization formalism, the $s$-wave BCS state can be written as
\begin{align}\label{eq:s_wave_2nd_quant}
    \vert \Psi_{\mathrm{BCS}} \rangle
    &= \prod_{\mathbf{k}} \Bigl( 1 + g_{\mathbf{k}} c^{\dagger}_{\mathbf{k},\uparrow} d^{\dagger}_{\mathbf{k},\downarrow} \Bigr) \vert 0 \rangle,
\end{align}
where $c^{\dagger}_{\mathbf{k},\uparrow}$ ($d^{\dagger}_{\mathbf{k},\downarrow}$) creates an electron (hole) in the top (bottom) layer and $\vert 0 \rangle$ corresponds to the vacuum state with the top layer empty and the LLL of bottom layer fully filled, i.e. all electrons are in the bottom layer. Since the number of electron-hole pairs must be definite in our calculations, we need to project Eq.~\eqref{eq:s_wave_2nd_quant} to the sector with pseudo-spin $S_z = 0$, i.e. the balanced bilayer case. The trial state then becomes
\begin{align}\label{eq:s_wave_2nd_quant_Sz=0}
    \vert \Psi_{\mathrm{BCS}} \rangle
    &= \sum_{ \{ g_{\mathbf{k}} \} } \prod_{\mathbf{k}} \Bigl( g_{\mathbf{k}} c^{\dagger}_{\mathbf{k},\uparrow} d^{\dagger}_{\mathbf{k},\downarrow} \Bigr) \vert 0 \rangle,
\end{align}
where the sum runs over all possible sets $\{ g_{\mathbf{k}} \}$ with $\vert \{ g_{\mathbf{k}} \} \vert = N_1$. The occupation number of orbital with momentum $\mathbf{k}$ is then given by the expectation value of number operators $c^{\dagger}_{\mathbf{k}}c_{\mathbf{k}}$ and $d^{\dagger}_{\mathbf{k}}d_{\mathbf{k}}$
\begin{equation}\label{eq:occ_extended}
    \begin{split}
        n_{\mathbf{k}}(\Delta)
        &= \langle c^{\dagger}_{\mathbf{k},\uparrow} c_{\mathbf{k},\uparrow} \rangle
        = \langle d^{\dagger}_{\mathbf{k},\downarrow} d_{\mathbf{k},\downarrow} \rangle \\
        & = \frac{1}{2} \biggl( 1 - \frac{ (k \ell_B)^2 - 1 }{\sqrt{ ((k \ell_B)^2 - 1)^2 + \Delta^2 }} \biggr).
    \end{split}
\end{equation}
The second line is the same as Eq.~\eqref{eq:occ} but written in terms of dimensionless parameters $k \ell_B$ and $\Delta$ (defined as the BCS order parameter in units of the Fermi energy), where $\ell_B = 1 / k_F$ is the magnetic length. Another useful parameter is the probability $\mathcal{P}(\mathbf{k})$ that a CF is in the orbital with momentum $\mathbf{k}$ (see Ref.~\cite{Simon1})
\begin{equation} \label{eq:occ_prob}
    \mathcal{P}(\mathbf{k}) = \frac{1}{2 N_1} \frac{ \partial \log \langle \tilde{\Psi}_{\mathrm{BCS}} (\{ g_{\mathbf{k}} \}) \vert \tilde{\Psi}_{\mathrm{BCS}} (\{ g_{\mathbf{k}} \}) \rangle }{ \partial \log g_{\mathbf{k}} },
\end{equation}
with $\tilde{\Psi}_{\mathrm{BCS}} (\{ g_{\mathbf{k}} \})$ as given in Eq.~\eqref{eq:s-wave_BCS}. In practice it is more convenient to use the LL indices $n$ as effective momentum labels. One can switch between them via
\begin{equation} \label{eq:k_n_relation}
    k \ell_B = \sqrt{ \frac{2}{2 N_1 - 1} \Bigl(n + \frac{1}{2}\Bigr) \Bigl(n + \frac{3}{2}\Bigr) }.
\end{equation}
Therefore the energy 
\begin{equation}
\label{eq:epsilon_n}
    \varepsilon_n=(k \ell_B)^2 - 1=\frac{2\Bigl(n + \frac{1}{2}\Bigr) \Bigl(n + \frac{3}{2}\Bigr)}{2 N_1 - 1}  -1.
\end{equation}
The occupation probabilities of a given CF orbital are given by
\begin{equation}\label{eq:nk_pk_relation}
    p_n  = \frac{N_1}{2 (n+1)} \mathcal{P}(\mathbf{k}) 
\end{equation}
The fact that we can compute the occupation probabilities with either $\{g_n\}$ (via Eq.~\eqref{eq:occ_prob} and \eqref{eq:k_n_relation}) or $\Delta$ (via Eq.~\eqref{eq:occ_extended}) independently motivates us to consider the dimensionless BCS order parameter as an alternative variational parameter. To do that, we minimize
\begin{equation} \label{eq:delta_opt_tbmin}
    f (\{g_n\}) = \sum_{n=0}^{N_{\mathrm{LL}}-1} \vert p_n (\{g_n\}) - n_{\mathbf{k}} (\Delta) \vert^2
\end{equation}
for each $\Delta$ in a chosen set. Note that $\Delta$ has no dependence on LL index $n$. Assuming states with the same occupation numbers are (for our purpose) equivalent, we can represent any BCS state parameterized by $\Delta$, which we call ``$\Delta$-states", with a BCS state parameterized by the set of $g_n$ parameters that minimizes $f ( \{ g_n \} )$. This then allows us to evaluate quantities such as state overlaps for $\Delta$-states the same way we do $\Psi_{\mathrm{BCS}} (\{g_n\})$'s.

\section{Monte Carlo Procedures}

We use the Monte Carlo (MC) methods to calculate the (high-dimensional) integrals. The 111 state is used universally as the sampling distribution function. As an example of the implementation procedure, we start with a random configuration of $\{ \Omega^{\uparrow} \}$ and $\{ \Omega^{\downarrow} \}$ corresponding to electron coordinates in the top and bottom layer, respectively. At each MC step, a new configuration is proposed by moving one randomly-chosen electron to a different position, and is accepted with probability given by the Metropolis algorithm
\begin{equation}
    \rho ( \{ \Omega^{\uparrow} \}, \{ \Omega^{\downarrow} \} ) = \vert \Psi_{111} ( \{ \Omega^{\uparrow} \}, \{ \Omega^{\downarrow} \} ) \vert^2,
\end{equation}
where $\Psi_{111}$ is normalized. Such a step is repeated until convergence.

To compute, for example, the overlap between the 111 state and the trial BCS state with $N_{\mathrm{MC}}$ samples, we use
\begin{equation}
    \begin{split}
        \langle \Psi_{\mathrm{BCS}} \vert \Psi_{111} \rangle
        & = \int d\Omega^{\uparrow}_1 \cdots d\Omega^{\uparrow}_{N_{\uparrow}} d\Omega^{\downarrow}_1 \cdots d\Omega^{\downarrow}_{N_{\downarrow}} \Psi_{\mathrm{BCS}}^{*} ( \{ \Omega^{\uparrow} \}, \{ \Omega^{\downarrow} \} ) \Psi_{111} ( \{ \Omega^{\uparrow} \}, \{ \Omega^{\downarrow} \} ) \\
        & = \frac{1}{N_{\mathrm{MC}}} \sum_{I=0}^{N_{\mathrm{MC}}-1} \frac{ \Psi_{\mathrm{BCS}}^{*} ( \{ \Omega^{\uparrow} \}_I, \{ \Omega^{\downarrow} \}_I ) \Psi_{111} ( \{ \Omega^{\uparrow} \}_I, \{ \Omega^{\downarrow} \}_I ) }{ \rho ( \{ \Omega^{\uparrow} \}_I, \{ \Omega^{\downarrow} \}_I ) } j ( \{ \Omega^{\uparrow} \}_I, \{ \Omega^{\downarrow} \}_I ),
    \end{split}
\end{equation}
where
\begin{equation}
    j ( \{ \Omega^{\uparrow} \}, \{ \Omega^{\downarrow} \} ) = \prod_{n=0}^{N_{\uparrow}-1} \sin \theta_n^{\uparrow} \prod_{n=0}^{N_{\downarrow}-1} \sin \theta_n^{\downarrow}
\end{equation}
is the Jacobian of unit surface area in spherical coordinates. $I$ is the label of a single MC sample.

\section{Hamiltonian}

We use the ``minimal" model Hamiltonian and consider only Coulomb interactions in a bilayer system
\begin{equation} \label{eq:V_split}
    \begin{split}
        V_{\uparrow \uparrow} (r)
        &= V_{\downarrow \downarrow} (r)
        = e^2 / ( 4 \pi \epsilon r ) \\
        V_{\uparrow \downarrow} (r)
        &= -e^2 / \Bigl( 4 \pi \epsilon \sqrt{ r^2 + d^2 } \Bigr),
    \end{split}
\end{equation}
where $d$ is the interlayer separation. The first and second line describes the intra- and inter-layer interactions, respectively. We work in the spherical geometry, where the particles are on a sphere of radius $R = \sqrt{q} \ell_B$ with $q = N_{\phi} / 2$ the half-flux number. The explicit form we use is (dropping an overall factor of $e^2/(4 \pi \epsilon)$)
\begin{equation} \label{eq:V_explicit}
    V_{\sigma_i \sigma_j} = \frac{1}{2} \sum_{i, j}^{N} \frac{ (-1)^{(1 + \delta_{\sigma_i \sigma_j})} }{\sqrt{(2R | u_i v_j - u_j v_i |)^2 + (1 - \delta_{\sigma_i \sigma_j}) d^2}},
\end{equation}
where $i$ and $j$ run over all particles in both layers and $\sigma_i, \sigma_j \in \{ \uparrow, \downarrow \} $ label the layer to which particle $i$ and $j$ belong, respectively. $u_i$ and $v_j$ are the spinor coordinates as defined in Eq.~\eqref{eq:spinor_coordinates}. Notice both inter- and intra-layer interactions are covered in this expression and the relative minus sign suggests that they compete with each other. Such competition is tuned by the layer separation $d$ and gives rise to the BEC-BCS crossover in quantum Hall bilayers.

When maximizing the overlap of $\alpha$-ansatz with the energy-minimized BCS ground states, a discontinuous jump in $\alpha_{\mathrm{opt}}$ can be observed at $d \approx 0.4\ell_B$, as seen in Fig.~\ref{fig:fig3}(b). This suggests the existence of a subspace (for a 6+6 system it is the range $1.5 \lesssim \alpha < \infty$) within the variational space spanned by $\alpha$-ansatz that is not accessible to the conventional BCS theory. To further investigate this ``non-physical" subspace, we generalize our Hamiltonian by introducing a non-negative ``weight" factor $\lambda$ 
\begin{equation} 
\label{eq:H_lambda}
    H_{\lambda} = \sum^{N}_{ i < j, \sigma } \frac{1}{| r_{i,\sigma} - r_{j,\sigma} |} - \lambda \sum_{i, j} \frac{1}{|r_{i,\uparrow} - r_{j,\downarrow} |}
\end{equation}
for the intralayer Coulomb interaction. Here we have used $\ell_B$ and $e^2 / \ell_B$ as the unit of length and energy, respectively. $\sigma \in \{\uparrow, \downarrow\}$ is again the layer label. For $0 \le \lambda \le 1$ we recover the model Hamiltonian and $\lambda$ is merely a trivial substitute for $d$. We identify $\lambda = 0$ as the CFL-limit ($d \to \infty$) where the two layers are completely decoupled and interlayer interaction is negligible, and $\lambda = 1$ as the 111-limit ($d = 0$) where intra- and inter-layer interactions are equally important. For $\lambda > 1$, however, the interlayer interaction is weighted more than its intralayer counterpart, which is outside the physically relevant regime for bilayers. As shown in Fig.~\ref{fig:alpha_opt_H_lambda}, the $\lambda$-factor grants us access to this non-physical subspace of the $\alpha$-ansatz, where the optimum value of $\alpha$ lies between $\sim 1$ and $\infty$.

\section{Supplementary figures}

\begin{figure}[htb]
    \centering
    \includegraphics[width=1.0\linewidth]{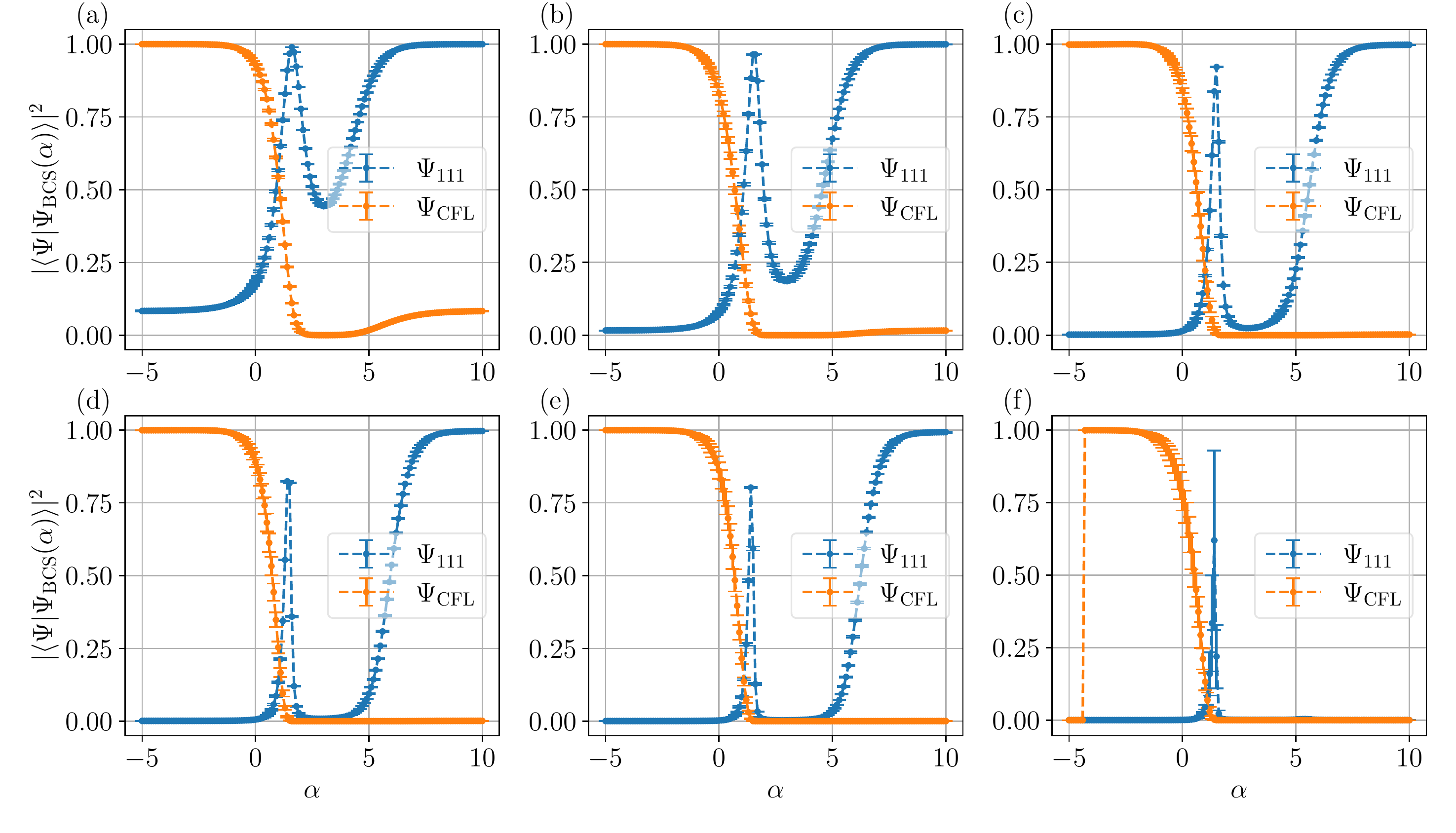}
    \caption{
        Overlaps of $\alpha$-state with the two model states, CFL- and 111-state, for different system sizes: (a) 4+4, (b) 5+5, (c) 7+7, (d) 8+8, (e) 9+9, and (f) 10+10. The $\alpha \to \infty$ state has very high overlap with the 111-state for all system sizes up to 9+9 electrons, whereas for a system of 10+10 electrons the $\alpha \sim 1$ state appears to have a higher overlap. However, the huge error bar at $\alpha \sim 1$ and the vanishing CFL-overlaps for $\alpha \lesssim -4.5$ are both clear indications of some form of numerical instability associated with larger systems. Note that the 111-overlap for $\alpha \sim 1$ state decreases as a function of system sizes.
    }
    \label{fig:model_ovlps_arr}
\end{figure}

\begin{figure}[htb]
    \centering
    \includegraphics[width=0.75\linewidth]{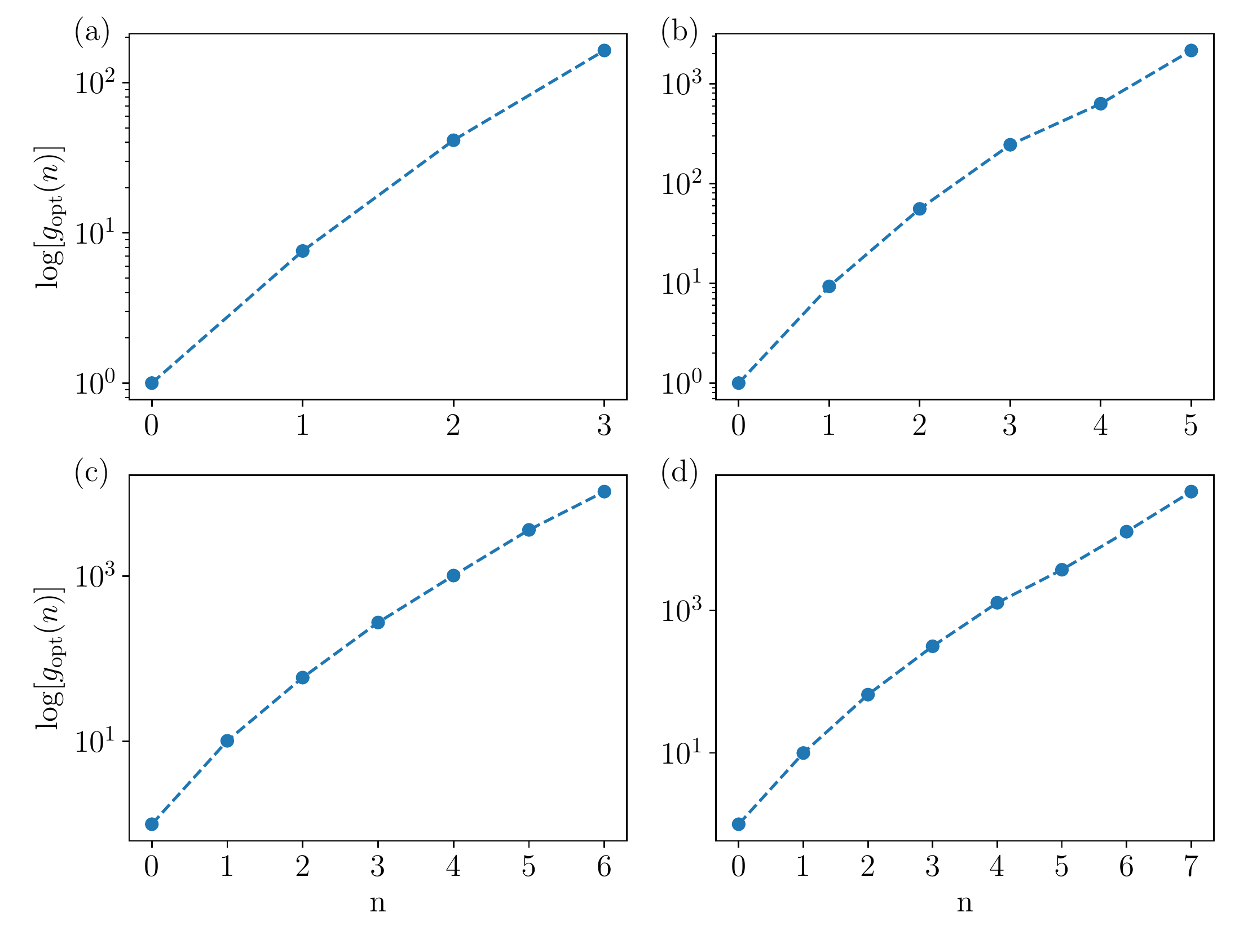}
    \caption{
        Values of optimum $g_n$ parameters (in logarithmic scale) corresponding to the variationally energy-minimized BCS state at small interlayer separation (i.e. the 111-limit) for different system sizes: (a) 4+4, (b) 6+6, (c) 7+7, and (d) 8+8. The exhibited linearity in a semi-log plot for all system sizes (up to 8+8) motivates the $\alpha$-ansatz Eq.~\eqref{eq: alpha-ansatz}. For larger system sizes, however, the optimization algorithm generally has difficulty finding the energy minimum.
    }
    \label{fig:g_opt_arr}
\end{figure}

\begin{figure}[htb]
    \centering
    \includegraphics[width=1.0\linewidth]{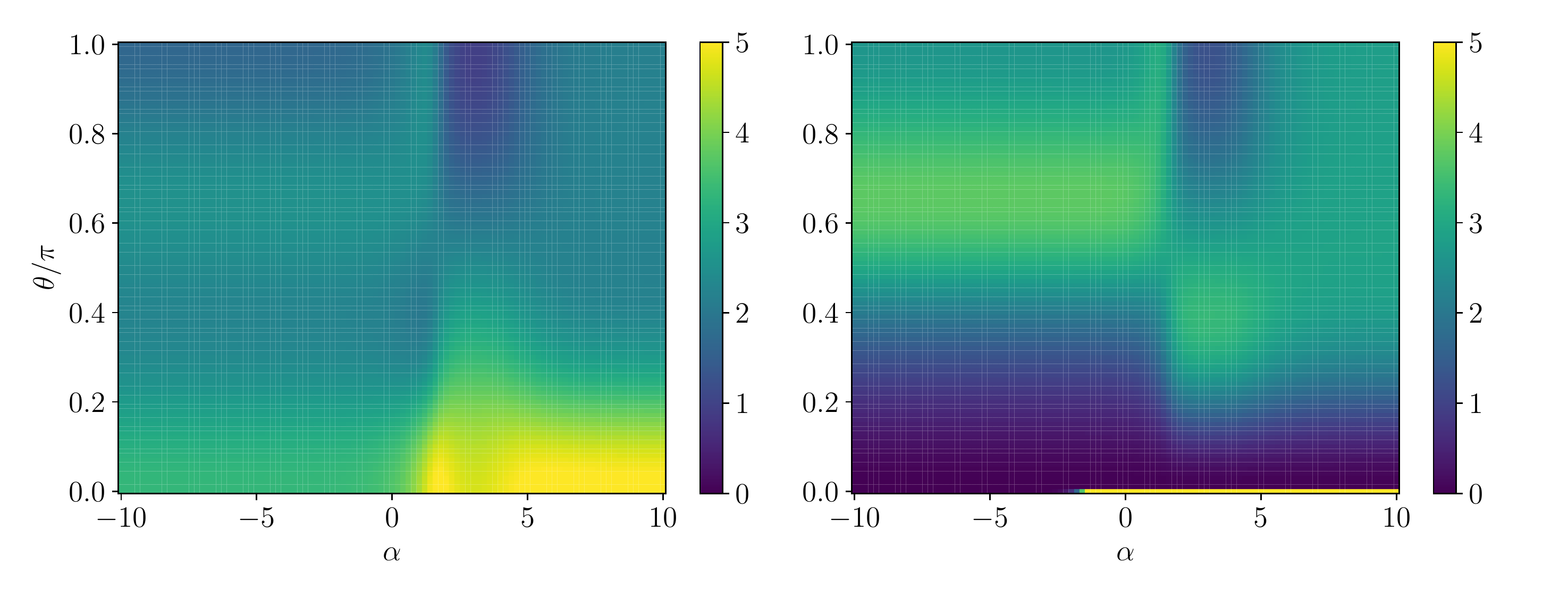}
    \caption{
        Pair correlations between interlayer (left) and intralayer (right) particles for $\alpha$-ansatz with $-10 \leq \alpha \leq 10$ in a system of 4+4 electrons. $\theta$ is (in spherical coordinates) the polar angle difference between the pair of particles. Since we have CFs in one layer and anti-CFs in the other, for small $\theta$ (i.e. when two particles are close together) the interlayer correlation displays a maximum for large $\alpha$ (pairs of CFs and anti-CFs form bosons and they bunch together) while the intralayer correlation shows a minimum (CFs are fermions and they avoid each other). The divergences in intralayer correlation at zero-$\theta$ are to be considered as numerical errors.
    }
    \label{fig:colorcorr_8_4}
\end{figure}

\begin{figure}[htb]
    \centering
    \includegraphics[width=1.0\linewidth]{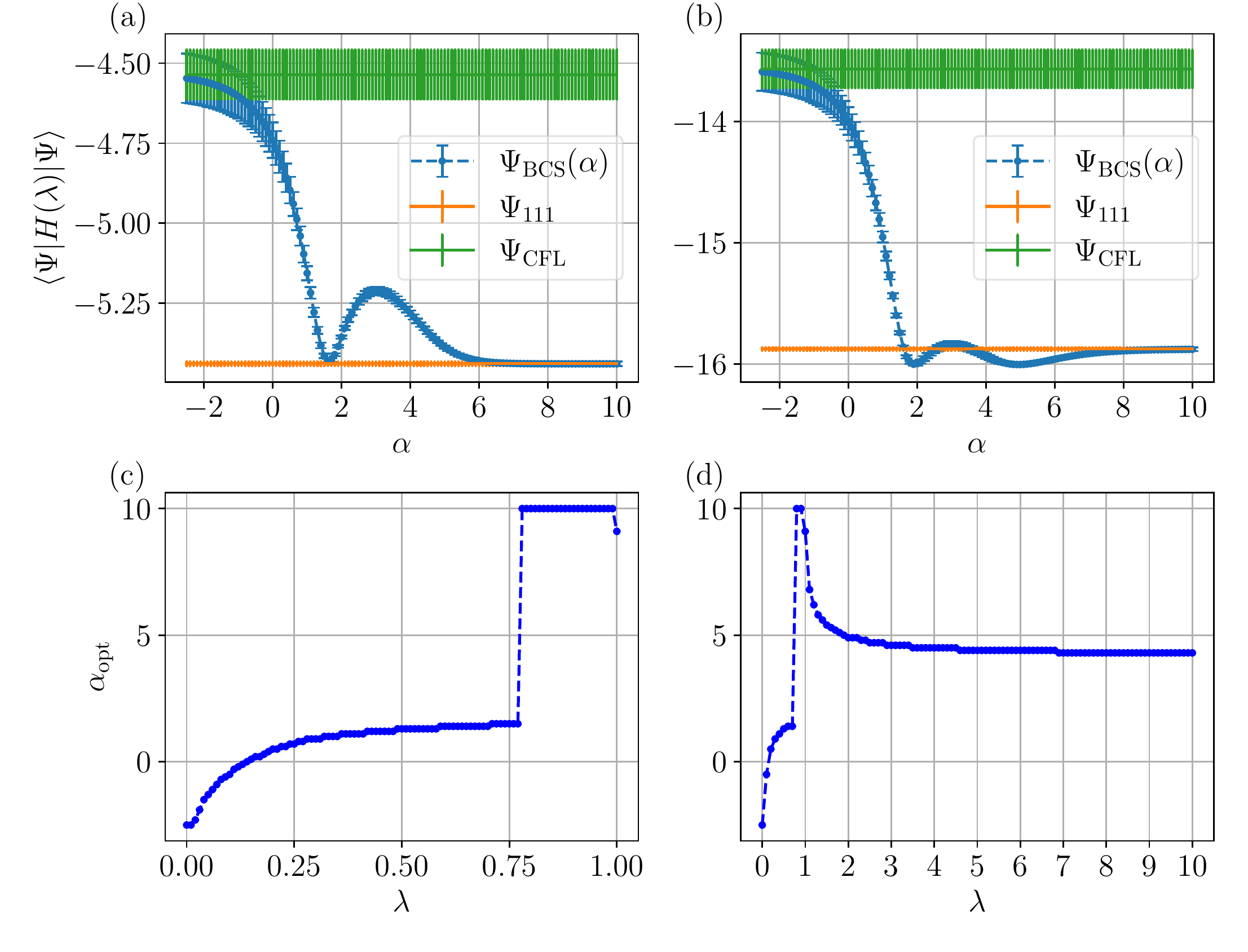}
    \caption{
        Top row: ($\lambda$-weighted) energy comparison between the two model states and the $\alpha$-ansatz with $-2.5 \leq \alpha \leq 10$ for (a) $\lambda = 1$ (the 111-limit) and (b) $\lambda = 2$ in the non-physical range. All energy expectation values are evaluated using the ``$\lambda$-Hamiltonian" Eq.~\eqref{eq:H_lambda}. For $\lambda = 1$, we have one local energy minimum at $\alpha \sim 1$ and approach the global minimum, i.e. energy of the 111 state (marked by the lower horizontal line), as $\alpha \to \infty$. This is in accordance with what is shown in Fig.~\ref{fig:alpha_opt}(a) where these two states have the highest overlaps with the 111 state. For $\lambda = 2$, BCS states with intermediate values of $\alpha$ appear to have even lower energies than the reference line, but the energy of the $\alpha \to \infty$ state still converges to that of the 111 state.
        Bottom row: optimum value of $\alpha$ corresponding to the energy minimum, $\langle \Psi_{\alpha} \vert H(\lambda) \vert \Psi_{\alpha} \rangle_{\mathrm{min}}$, for (c) $0 \leq \lambda \leq 1$ (physical range) and (d) $0 \leq \lambda \leq 10$. The discontinuous jump in $\alpha_{\mathrm{opt}}$ at $\alpha \sim 0.76$ corresponds to the discontinuity at $d / \ell_B \sim 0.4$ as seen in Fig.~\ref{fig:fig3}(b), leaving behind a ``forbidden" subspace in the Hilbert space of $\alpha$-ansatz that is only accessible by going beyond the physically-relevant regime, i.e. by setting $\lambda > 1$. All results above are for a bilayer system of $4+4$ electrons. With the $H_{\lambda}$ approach, we have shown that energy minimization and overlap maximization are essentially equivalent methods for finding the optimum $\alpha$-ansatz.
    }
    \label{fig:alpha_opt_H_lambda}
\end{figure}

\end{appendix}
\end{document}